\documentclass[twocolumn,showpacs,preprintnumbers,amsmath,amssymb]{revtex4}
\usepackage{graphics}% Include figure files
\usepackage{graphicx}% Include figure files
\usepackage{dcolumn}% Align table columns on decimal point
\usepackage{bm}% bold math
\usepackage{epsfig}

\def\reff{\vskip 4pt \par \hangindent 16pt \noindent}

\begin{document}

%***************
\begin{flushright} 
% gr-qc/0410044 
\end{flushright} 

\title{Experimental Design for the LATOR Mission}

\author{{Slava G. Turyshev},\footnote{Electronic address:
        turyshev@jpl.nasa.gov}$^a$  
{Michael Shao},\footnote{Electronic address: 
        mshao@huey.jpl.nasa.gov}$^a$ and  
{Kenneth Nordtvedt, Jr.}\footnote{Electronic address: kennordtvedt@imt.net}$^b$ \\
~~~~ }

\affiliation{$^a$Jet Propulsion Laboratory, California Institute
of  Technology, Pasadena, CA 91109 \\
$^b$Northwest Analysis, 118 Sourdough Ridge Road, Bozeman MT 59715 USA}

%***********************************************
% \date{\today}
%**************************************************
%
%%************************ABSTRACT

\begin{abstract}
This paper discusses experimental design for the Laser Astrometric Test Of Relativity (LATOR) mission. LATOR is designed to reach unprecedented accuracy of 1 part in 10$^{8}$ in measuring the curvature of the solar gravitational field as given by the value of the key Eddington post-Newtonian parameter $\gamma$.  This mission will demonstrate the accuracy needed to measure effects of the next post-Newtonian order ($\propto G^2$) of light deflection resulting from gravity's intrinsic non-linearity.  LATOR will provide the first precise measurement of the solar quadrupole moment parameter, $J_2$, and will improve determination of a variety of relativistic effects including Lense-Thirring precession.  The mission will benefit from the recent progress in the optical communication technologies -- the immediate and natural step above the standard radio-metric techniques. The key element of LATOR is a geometric redundancy provided by the laser ranging and long-baseline optical interferometry.  We discuss the mission and optical designs, as well as the expected performance of this proposed mission. LATOR will lead to very robust advances in the tests of Fundamental physics: this mission could discover a violation or extension of general relativity, or reveal the presence of an additional long range interaction in the physical law.  There are no analogs to the LATOR experiment; it is unique and is a natural culmination of solar system gravity experiments.
\end{abstract}


\maketitle

%********************* CONTENTS 
 
%\tableofcontents
%\listoffigures
%\listoftables
 
%*********************1) INTRODUCTION
\section{Introduction}
\label{sec:intro}

After almost ninety years since general relativity was born, Einstein's theory has survived every test. Such a longevity, along with the absence of any adjustable parameters, does not mean that this theory is absolutely correct, but it serves to motivate more accurate tests to determine the level of accuracy at which it is violated. 
Einstein's general theory of relativity (GR) began with its empirical success in 1915 by explaining the anomalous perihelion precession of Mercury's orbit, using no adjustable theoretical parameters.  Shortly thereafter, Eddington's 1919 observations of star lines-of-sight during a solar eclipse confirmed the doubling of the deflection angles predicted by GR as compared to Newtonian-like and Equivalence Principle arguments.  This conformation made the general theory of relativity an instant success. 

From these beginnings, the general theory of relativity has been verified at ever higher accuracy. Thus, microwave ranging to the Viking Lander on Mars yielded accuracy  $\sim$0.2\% in the tests of GR \cite{viking_shapiro1,viking_reasen,viking_shapiro2}. Spacecraft and planetary radar observations reached an accuracy of $\sim$0.15\% \cite{anderson02}.  The astrometric observations of quasars on the solar background performed with Very-Long Baseline Interferometry (VLBI) improved the accuracy of the tests of gravity to $\sim$0.045\% \cite{RoberstonCarter91,Lebach95,Shapiro_SS_etal_2004}. Lunar laser ranging,  a continuing legacy of the Apollo program, provided $\sim$0.011\% verification of GR via precision measurements of the lunar orbit \cite{Ken_LLR68,Ken_LLR91,Ken_LLR30years99,Ken_LLR_PPNprobe03,JimSkipJean96,Williams_etal_2001,LLR_beta_2004}. Finally, the recent experiments with the Cassini spacecraft improved the accuracy of the tests to $\sim$0.0023\% \cite{cassini_ber}. As a result general relativity became the standard theory of gravity when astrometry and spacecraft navigation are concerned. 

However, the tensor-scalar theories of gravity, where the usual general relativity tensor field coexists with one or several long-range scalar fields, are believed to be the most promising extension of the theoretical foundation of modern gravitational theory. The superstring, many-dimensional Kaluza-Klein, and inflationary cosmology theories have revived interest in the so-called `dilaton fields', i.e. neutral scalar fields whose background values determine the strength of the coupling constants in the effective four-dimensional theory. The importance of such theories is that they provide a possible route to the quantization of gravity and unification of physical law. 

Recent theoretical findings suggest that the present agreement between Einstein's theory and experiment might be naturally compatible with the existence of a scalar contribution to gravity. In particular, Damour and Nordtvedt \cite{damour_nordtvedt} (see also 
\cite{DamourPolyakov94} for non-metric versions of this mechanism and \cite{DPV02} for the recent summary of a dilaton-runaway scenario) have found that a scalar-tensor theory of gravity may contain a `built-in' cosmological attractor mechanism towards GR.  A possible scenario for cosmological evolution of the scalar field was given in \cite{Ken_LLR_PPNprobe03,damour_nordtvedt}. Their speculation assumes that the parameter  $\frac{1}{2}(1-\gamma)$  was of order of 1 in the early universe, at the time of inflation, and has evolved to be close to, but not exactly equal to, zero at the present time. In fact, the analyzes discussed above not only motivate new searches for very small deviations of relativistic gravity in the solar system, they also predict that such deviations are currently present in the range from 10$^{-5}$ to $\sim 5\times 10^{-8}$ of the post-Newtonian effects.  This would require measurement of the effects of the next post-Newtonian order ($\propto G^2$) of light deflection resulting from gravity's intrinsic non-linearity. An ability to measure the first order light deflection term at the accuracy comparable with the effects of the second order is of the utmost importance for the gravitational theory and is the challenge for the 21st century fundamental physics. 

The Eddington parameter $\gamma$, whose value in general relativity is unity, is perhaps the most fundamental PPN parameter, in that $(1-\gamma)$ is a measure, for example, of the fractional strength of the scalar gravity interaction in scalar-tensor theories of gravity \cite{Damour_EFarese96}.  Within perturbation theory for such theories, all other PPN parameters to all relativistic orders collapse to their general relativistic values in proportion to $(1-\gamma)$. This is why measurement of the first order light deflection effect at the level of accuracy comparable with the second-order contribution would provide the crucial information separating alternative scalar-tensor theories of gravity from GR \cite{Ken_2PPN_87} and also to probe possible ways for gravity quantization and to test modern theories of cosmological evolution \cite{damour_nordtvedt,DamourPolyakov94,DPV02}. 

The LATOR mission is designed to directly address the issue above with an unprecedented accuracy \cite{lator_cqg_2004}. The test will be performed in the solar gravity field using optical interferometry between two micro-spacecraft.  Precise measurements of the angular position of the spacecraft will be made using a fiber coupled multi-chanelled optical interferometer on the ISS with a 100 m baseline. The primary objective of the LATOR mission will be to measure the gravitational deflection of light by the solar gravity to accuracy of 0.1 picoradians (prad) ($\sim0.02 ~\mu$as), which corresponds to $\sim$10 picometers (pm) on a 100 m interferometric baseline.

A combination of laser ranging among the spacecraft and direct interferometric measurements will allow LATOR to measure deflection of light in the solar gravity by a factor of more than 3,000 better than had recently been accomplished with the Cassini spacecraft. In particular, this mission will not only measure the key PPN parameter $\gamma$ to unprecedented levels of accuracy of one part in 10$^8$. As a result, this experiment will measure values of other PPN parameters such as parameter $\delta$ to 1 part in $10^3$ and discussion thereafter), the solar quadrupole moment parameter $J_2$ to 1 part in 20, and the frame dragging effects on light due to the solar angular momentum to precision of 1 parts in $10^2$.

The LATOR mission technologically is a very sound concept; all technologies that are needed for its success have been already demonstrated as a part of the JPL's Space Interferometry Mission (SIM) development. Technology that has become available in the last several years such as low cost microspacecraft, medium power highly efficient solid state and fiber lasers, and the development of long range interferometric techniques make possible an unprecedented factor of 3,000 improvement in this test of general relativity possible. This mission is unique and is the natural next step in solar system gravity experiments which fully exploits modern technologies.

This paper organized as follows:  Section \ref{sec:lator_description} provides the overview for the LATOR experiment including the preliminary mission design. In Section \ref{sec:lator_current} we discuss the current design for the LATOR flight system. In Section \ref{sec:error_bud} we will discuss the expected performance for the LATOR instrument. 
Section \ref{sec:conc} discusses the next steps that will taken in the development of the LATOR mission.  

\section{Overview of LATOR}
\label{sec:lator_description}

\begin{figure*}[t!]
 \begin{center}
\noindent    
\psfig{figure=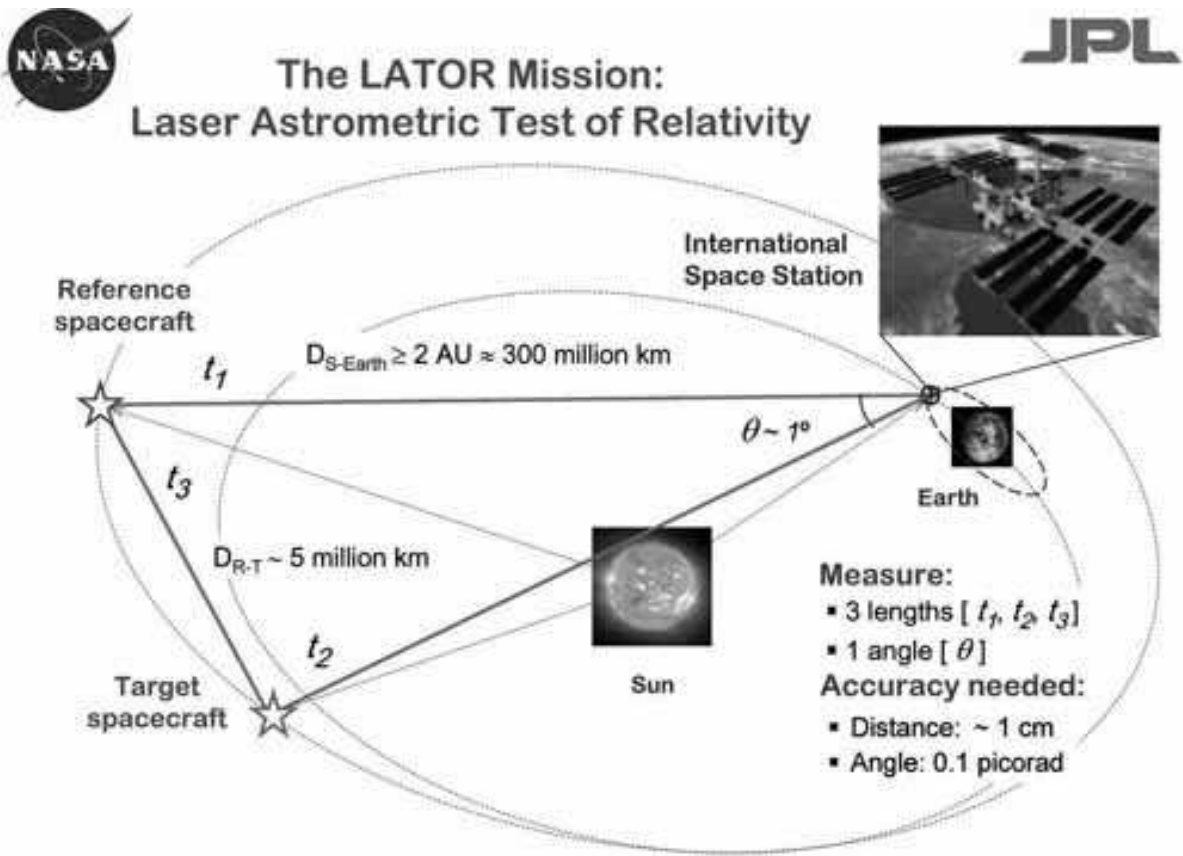,width=158mm}%,height=90mm}
\end{center}
\vskip -10pt 
  \caption{The overall geometry of the LATOR experiment.  
 \label{fig:lator}}
\end{figure*}

The LATOR experiment uses laser interferometry between two micro-spacecraft whose lines of sight pass close by the Sun to accurately measure deflection of light in the solar gravity \cite{lator_cqg_2004}. Another component of the experimental design is a long-baseline stellar optical interferometer placed on the ISS. Figure \ref{fig:lator} shows the general concept for the LATOR missions including the mission-related geometry, experiment details  and required accuracies. 

We shall now discuss the LATOR mission in detail.

%************

%**************
\subsection{Mission Design}

As evident from Figure \ref{fig:lator}, the key element of the LATOR experiment is a redundant geometry optical truss to measure the departure from Euclidean geometry caused by gravity.  Two spacecraft are injected into a heliocentric solar orbit on the opposite side of the Sun from the Earth. The triangle in figure has three independent quantities but three arms are monitored with laser metrology. In particular, each spacecraft equipped with a laser ranging system that enable a measurement of the arms of the triangle formed by the two spacecraft and the ISS.   According to Euclidean rules this determines a specific angle at the interferometer; LATOR can  directly measure this angle directly. In particular, the laser beams transmitted by each spacecraft are detected by a long baseline ($\sim$ 100 m) optical interferometer on the ISS. The actual angle measured at interferometer is compared to angle calculated using Euclidean rules and three side measurements; the difference is the non-Euclidean deflection signal [which varies in time during spacecraft passages] which contains the scientific information.
Therefore, the uniqueness of this mission comes with its built-in geometrically redundant architecture that enables LATOR to measure the departure from Euclidean geometry caused by the solar gravity field to a very high accuracy. The accurate measurement of this departure constitutes the primary mission objective.

To enable the primary objective, LATOR will place two spacecraft into a heliocentric orbit so that observations may be made when the spacecraft are behind the Sun as viewed from the ISS.   The two spacecraft are to be separated by about 1$^\circ$, as viewed from the ISS \cite{footnote,yu94}. With the help of the JPL Advanced Project Design Team (Team X), we recently conducted a detailed mission design studies \cite{teamx}. In particular, we analyzed various trajectory options for the deep-space flight segment of LATOR, using both Orbit Determination Program (ODP) and Satellite Orbit Analysis Program (SOAP) --  the two standard JPL navigation software packages. 

An orbit with a 3:2 resonance with the Earth was found to uniquely satisfy the LATOR orbital requirements \cite{teamx}. (The 3:2 resonance occurs when the Earth does 3 revolutions around the Sun while the spacecraft does exactly 2 revolutions of a 1.5 year orbit. The exact period of the orbit may vary slightly ($<$1\%) from a 3:2 resonance depending on the time of launch.) For this orbit, in 13 months after the launch, the spacecraft are within $\sim10^\circ$ of the Sun with first occultation occuring in 15 months after launch \cite{lator_cqg_2004}.  At this point, LATOR is orbiting at a slower speed than the Earth, but as LATOR approaches its perihelion, its motion in the sky begins to reverse and the spacecraft is again occulted by the Sun 18 months after launch.  As the spacecraft slows down and moves out toward aphelion, its motion in the sky reverses again and it is occulted by the Sun for the third and final time 21 months after launch.  

The 3:2 Earth resonant orbit provides an almost ideal trajectory for the LATOR mission, specifically i) it imposes no restrictions on the time of launch; ii) with a small propulsion maneuver after launch, it places the two LATOR spacecraft at the distance of less then 3.5$^\circ$ (or $\sim 14 ~R\odot$) for the entire duration of the experiment (or $\sim$8 months); iii) it provides three solar conjunctions even during the nominal mission lifetime of 22 months, all within 7 month period; iv) at a cost of an extra maneuver, it offers a possibility of achieving  small orbital inclinations (to enable measurements at different solar latitudes), and, finally, v) it offers a very slow change in the Sun-Earth-Probe (SEP) angle of about $\sim R_\odot$ in 4 days. As such, this orbit represents a very attractive choice for LATOR.

We intend to further study this 3:2 Earth resonant trajectory as the baseline option for the mission. In particular, there is an option to have the two spacecraft move in  opposite directions during the solar conjunctions. This option will increase the amount of $\Delta v$ LATOR should carry on-board, but it significantly reduces the experiment's dependence on the accuracy of determination of the solar impact parameter. This particular option is currently being investigated and results will be reported elsewhere. 

We shall now consider the basic elements of the LATOR optical design. 

\subsection{Optical Design}

\begin{figure*}[t!]
 \begin{center}
\noindent    
\psfig{figure=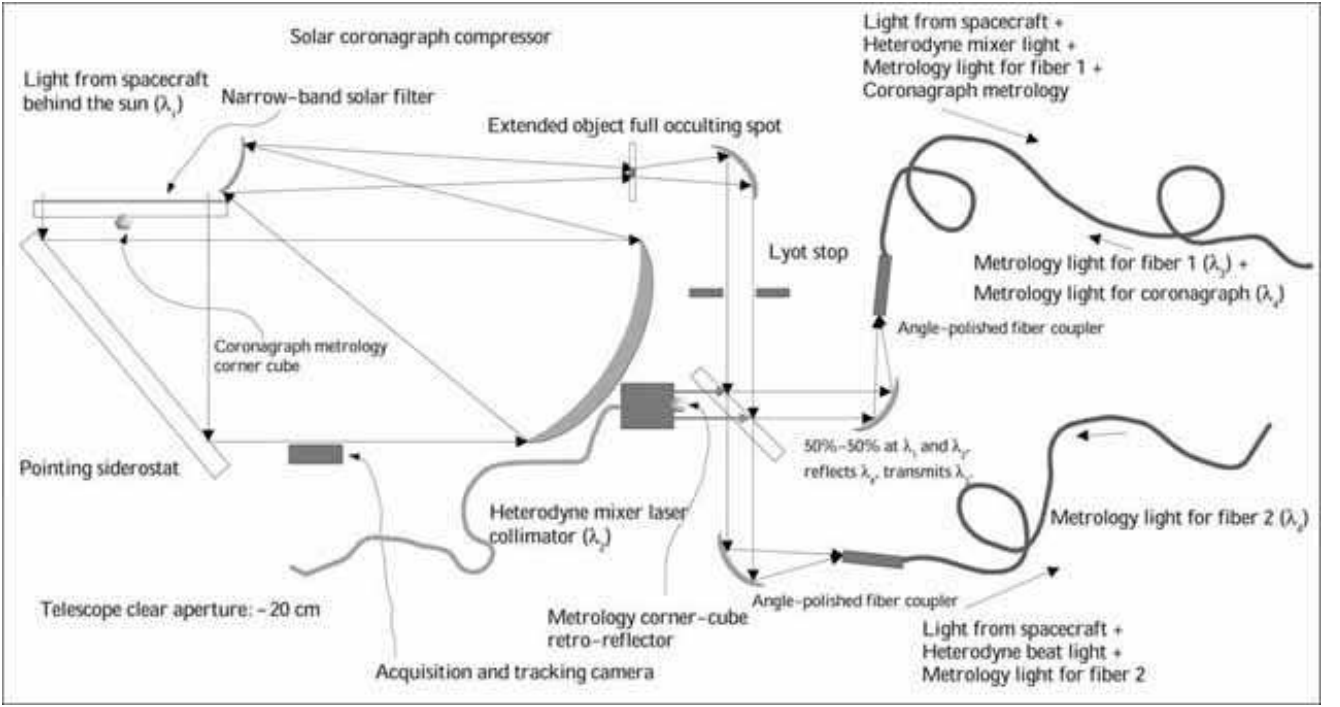,width=175mm}%,height=90mm}
\end{center}
\vskip -10pt 
  \caption{Basic elements of the LATOR optical design. The laser light (together with the solar background) is going through a full aperture ($\sim 20$cm) narrow band-pass filter with $\sim 10^{-4}$ suppression properties. The remaining light illuminates the baseline metrology corner cube and falls onto a steering flat mirror where it is reflected to an off-axis telescope with no central obscuration (needed for metrology). It is then enters the solar coronograph compressor by first going through a 1/2 plane focal plane occulter and then coming to a Lyot stop. At the Lyot stop, the background solar light is reduced by a factor of $10^{6}$. The combination of a narrow band-pass filter and coronograph enables the solar luminosity reduction from $V=-26$ to $V=4$ (as measured at the ISS), thus, enabling the LATOR precision observations.
\label{fig:optical_design}}
\end{figure*}

A single aperture of the interferometer on the ISS consists of three 20 cm diameter telescopes (see Figure \ref{fig:optical_design} for a conceptual design). One of the telescopes with a very narrow bandwidth laser line filter in front and with an InGAs camera at its focal plane, sensitive to the 1.3 $\mu$m laser light, serves as the acquisition telescope to locate the spacecraft near the Sun.

The second telescope emits the directing beacon to the spacecraft. Both spacecraft are served out of one telescope by a pair of piezo controlled mirrors placed on the focal plane. The properly collimated laser light ($\sim$10W) is injected into the telescope focal plane and deflected in the right direction by the piezo-actuated mirrors. 

The third telescope is the laser light tracking interferometer input aperture which can track both spacecraft at the same time. To eliminate beam walk on the critical elements of this telescope, two piezo-electric X-Y-Z stages are used to move two single-mode fiber tips on a spherical surface while maintaining focus and beam position on the fibers and other optics. Dithering at a few Hz is used to make the alignment to the fibers and the subsequent tracking of the two spacecraft completely automatic. The interferometric tracking telescopes are coupled together by a network of single-mode fibers whose relative length changes are measured internally by a heterodyne metrology system to an accuracy of less than 10 pm.

%************
%**************

The spacecraft  are identical in construction and contain a relatively high powered (1 W), stable (2 MHz per hour $\sim$  500 Hz per second), small cavity fiber-amplified laser at 1.3 $\mu$m. Three quarters of the power of this laser is pointed to the Earth through a 20 cm aperture telescope and its phase is tracked by the interferometer. With the available power and the beam divergence, there are enough photons to track the slowly drifting phase of the laser light. The remaining part of the laser power is diverted to another telescope, which points towards the other spacecraft. In addition to the two transmitting telescopes, each spacecraft has two receiving telescopes.  The receiving telescope on the ISS, which points towards the area near the Sun, has laser line filters and a simple knife-edge coronagraph to suppress the Sun light to 1 part in $10^4$ of the light level of the light received from the space station. The receiving telescope that points to the other spacecraft is free of the Sun light filter and the coronagraph.

In addition to the four telescopes they carry, the spacecraft also carry a tiny (2.5 cm) telescope with a CCD camera. This telescope is used to initially point the spacecraft directly towards the Sun so that their signal may be seen at the space station. One more of these small telescopes may also be installed at right angles to the first one to determine the spacecraft attitude using known, bright stars. The receiving telescope looking towards the other spacecraft may be used for this purpose part of the time, reducing hardware complexity. Star trackers with this construction have been demonstrated many years ago and they are readily available. A small RF transponder with an omni-directional antenna is also included in the instrument package to track the spacecraft while they are on their way to assume the orbital position needed for the experiment. 

The LATOR experiment has a number of advantages over techniques which use radio waves to measure gravitational light deflection. Advances in optical communications technology, allow low bandwidth telecommunications with the LATOR spacecraft without having to deploy high gain radio antennae needed to communicate through the solar corona. The use of the monochromatic light enables the observation of the spacecraft almost at the limb of the Sun, as seen from the ISS. The use of narrowband filters, coronagraph optics and heterodyne detection will suppress background light to a level where the solar background is no longer the dominant noise source. In addition, the short wavelength allows much more efficient links with smaller apertures, thereby eliminating the need for a deployable antenna. Finally, the use of the ISS will allow conducting the test above the Earth's atmosphere -- the major source of astrometric noise for any ground based interferometer. This fact justifies LATOR as a space mission.

\subsection{LATOR Interferometry}
\label{sec:heterodine}

In this section, we describe how angle measurements are made using the LATOR ground based interferometer. Since the spacecraft are monochromatic sources, the interferometer can efficiently use heterodyne detection to measure the phase of the incoming signal. The next section provides a simplified explanation of heterodyne interferometry for LATOR interferometer. This section also describes the use of ISS's orbit to resolve the fringe ambiguity that arises from using monochromatic signals.  

%%************
\begin{figure}[h!]
 \begin{center}
\noindent  \vskip -5pt   
\psfig{figure=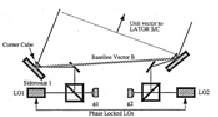,width=85mm}%,height=90mm}
\end{center}
\vskip -15pt 
  \caption{Heterodyne interferometry on 1 spacecraft with phase locked local oscillator.
 \label{fig:heterodyne1}}
%\end{figure} 
%%**************
%%%************
%\begin{figure}[h!]
 \begin{center}
\noindent  \vskip -5pt   
\psfig{figure=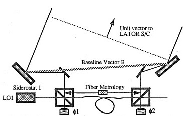,width=85mm}%,height=90mm}
\end{center}
\vskip -15pt 
  \caption{Fiber-linked heterodyne interferometry and fiber metrology system.
 \label{fig:heterodyne2}}
%\end{figure} 
%%**************
%%%************
%\begin{figure}[h!]
 \begin{center}
\noindent  \vskip -5pt   
\psfig{figure=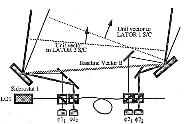,width=85mm}%,height=90mm}
\end{center}
\vskip -15pt 
  \caption{Heterodyne interferometry on 2 spacecraft.
 \label{fig:heterodyne3}}
\end{figure} 
%**************

\subsubsection{Heterodyne Interferometry}

Figures \ref{fig:heterodyne1}-\ref{fig:heterodyne3} show a simplified schematic of now angles are measured using a heterodyne interferometer. In Fig. \ref{fig:heterodyne1}, two siderostats are pointed at a target Two fiducials, shown as corner cubes, define the end points of the interferometer baseline . The light from each of the two arms is interfered with stable local oscillators (LOs) and the phase difference recorded. If the LOs in each arm were phase locked, the angles of the target with respect to the baseline normal is
{}
\begin{equation}
\theta=\arcsin[\frac{(2\pi n+\phi_1-\phi_2)\lambda}{2\pi b}]
\end{equation}

\noindent 
where $\lambda$ is the wavelength of the downlink laser, $n$ is an unknown integer arising from the fringe ambiguity and $b$ is the baseline length. In order to resolve this ambiguity multiple baselines were used in the previous mission design. This is discussed in greater detail in \cite{yu94}. In reality, it is difficult to phase lock the two LOs over the long baseline lengths. Figure \ref{fig:heterodyne2} shows how a single LO can be used and transmitted to both siderostats using a single mode fiber. In this configuration, a metrology system is used to monitor changes in the path length as seen by the LO as it propagates through the fiber. The metrology system measures the distance from one beam sputter to the other. In this case, the angle is given by
\begin{equation}
\theta=\arcsin[\frac{(2\pi n+\phi_1-\phi_2 +m_1)\lambda}{2\pi b}]
\end{equation}
\noindent
where $m_l$ is the phase variations introduced by changes in the optical path of the fiber as measured by the metrology system. 

Now consider the angle measurement between two spacecraft (Fig. \ref{fig:heterodyne3}). In this case the phase variations due to changes in the path through the fiber are common to both spacecraft. The differential angle is
 \begin{eqnarray}
\theta&=&\arcsin\big[\frac{(2\pi (n_1-n_2)+(m_1-m_2))\lambda}{2\pi b}+
\nonumber\\
&&\hskip 20pt +~
\frac{((\phi1_1-\phi1_2)-(\phi2_1-\phi2_2))\lambda}{2\pi b}\big]
\end{eqnarray}
\noindent
Since the spacecraft are monochromatic sources, the interferometer can efficiently use heterodyne detection to measure the phase of the incoming signal.  Note that because this is a differential measurement, it is independent of the any changes in the fiber length. In reality, the interferometer will have optical paths that are different between the two spacecraft signal paths. These paths that must be monitored accurately with a metrology system to correct for phase changes in the optical system due to thermal variations. However, this metrology must only measure path lengths in each ground station and not along the entire length of the fiber.  A more detailed design for the LATOR interferometer is given in Sec. \ref{sec:interf-ISS}.

\subsubsection{Resolving the Fringe Ambiguity}

The use of multiple interferometers is a standard solution to resolve the fringe ambiguity resulting from the interferometric detection of monochromatic light \cite{lator_cqg_2004}. 
The current LATOR mission proposal is immune for the fringe ambiguity problem as the orbit of the ISS provides enough variability (at least $\sim 30$\%) in the baseline projection. This variablity enables one to take multiple measurements during one orbit, in order to uniquely resolve the baseline orientation for each ISS orbit, making the fringe ambiguity not a problem for LATOR. 

\section{LATOR Flight System}
\label{sec:lator_current}

The LATOR flight system consists of two major components -- the deep-space component that will be used to transmit and receive the laser signals needed to make the science measurements and the interferometer on the ISS that will be used to interferometrically measure the angle between the two spacecraft and to transmit and receive the laser ranging signal to each of the spacecraft. 

In this Section we will discuss the design of these components in more details.

\subsection{LATOR Spacecraft}

\begin{figure}[h!]
 \begin{center}
\noindent  \vskip -5pt   
\psfig{figure=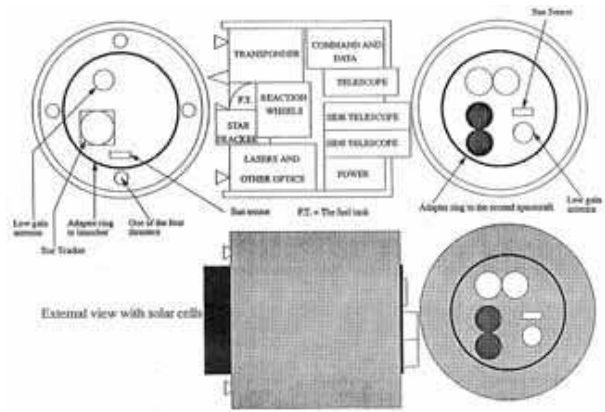,width=80mm}%,height=90mm}
\end{center}
\vskip -15pt 
  \caption{LATOR spacecraft concept.
 \label{fig:sc_concept}}
\end{figure}

There are two LATOR spacecraft in the deep-space component of the mission which will be used to transmit and receive the laser signals needed to make the science measurements. Figure \ref{fig:sc_concept} shows a schematic of the flight system without the solar cell array. The flight system is subdivided into the instrument payload and the spacecraft bus. The instrument includes the laser ranging and communications hardware and is described in more detail in the following section. The spacecraft contains the remainder of the flight hardware which includes solar cells, attitude control, and the spacecraft structure.

%%************
%**************

The LATOR spacecraft, like most spacecraft, will be composed of the following subsystems: thermal, structural, attitude control (ACS), power, command and data handling, telecommunications, and propulsion, in particular:

%************
\begin{figure}[!h!]
 \begin{center}
\noindent    
\psfig{figure=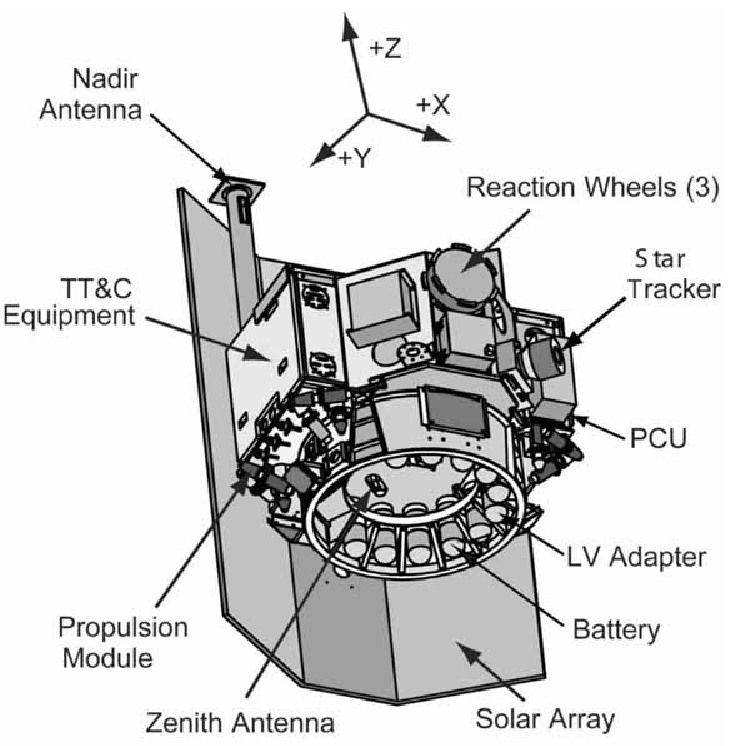,width=80mm}%,height=90mm}
\end{center}
\vskip -10pt 
% \caption{Typical configuration of a Spectrum Astro SA-200S/B bus.
% \label{fig:SA200S_config}}
%\end{figure} 
%%
%%%**************
%%************
%\begin{figure}[t!]
 \begin{center}
\noindent    
\psfig{figure=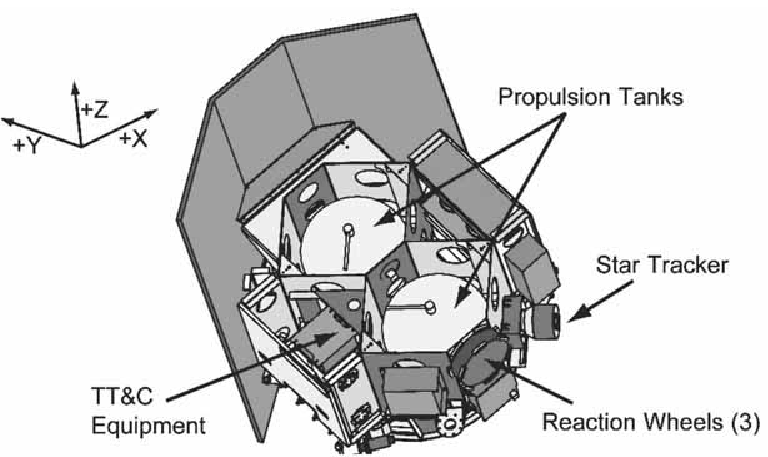,width=80mm}%,height=90mm}
\end{center}
\vskip -10pt 
  \caption{A typical Spectrum Astro SA-200S/B bus. With minor modifications this configuration may be adopted for the deep-space component of the LATOR mission.
 \label{fig:SA200S_config2}}
\end{figure} 
%**************

\begin{itemize}
\item {\bf Thermal:} The basic thermal design will similar to that of the SA-200B, with modifications to account for the variation in range. This design uses basically passive thermal control elements with electric heaters/thermostats.  The thermal control flight elements are multilayer insulation (MLI), thermal surfaces, thermal conduction control, and sensors.  The active elements are minimized and will be only electric heaters/thermostats.  To minimize heater power thermal louvers may be used. The current design assumes that spacecraft uses passive thermal control with heaters/thermostats because it is basically designed for Earth orbit.  

\item {\bf Structural:} The current best estimate (CBE) for the total dry mass is 115kg including a set of required modifications to the standard SA-200B bus (i.e. a small propulsion system, a 0.5m HGA for deep-space telecom, etc.)
The design calls for launching the two spacecraft side-by-side on a custom carrier structure as they should easily fit into the fairing (for instance, Delta II 2425-9.5). The total launch mass for the two spacecraft will be 552 kg.   This estimate may be further reduced, given more time to develop a point design. 

\item {\bf Attitude Control:} An attitude control system may be required to have pointing accuracy of 6 $\mu$rad and a pointing knowledge of 3 $\mu$rad. This may be achieved using a star tracker and sun sensor combination to determine attitude together with reaction wheels (RW) to control attitude. Cold-gas jets may be used to desaturate RWs. A Spectrum Astro SA-200B 3-axis stabilized bus with RWs for fine pointing and thrusters for RW desaturation is a good platform \cite{teamx}. 
For the current experiment design it is sufficient to utilize a pointing architecture with the following performance (3 sigma, per axis): control  6 $\mu$rad; knowledge 3 $\mu$rad; stability 0.1 $\mu$rad/sec.  The SA-200B readily accommodates these requirements.

\item {\bf Power Subsystem:} The flight system will require $\sim$50 W of power. This may be supplied by a 1 square meter GaAs solar cell array. To maintain a constant attitude with respect to the Sun, the solar cells must be deployed away from the body of the spacecraft. This will allow the cells to radiate away its heat to maintain the cells within their operating temperature range.

\item {\bf Telecommunications:} The telecommunications subsystem will be a hybrid which will utilize the optical communications capability of the instrument as the primary means of transmitting and receiving commands and data. In addition, a small low gain antenna for low data rate radio communications will be used for emergency purposes. This system will use a 15 W transmitter and 10 dB gain antenna. Using X band this system can operate with a 5 bit per second (bps) data rate. The design calls for an SDST X-Band transponder, operating at 15 W X-Band SSPA, a 0.5m HGA, two X-Band LGAs pointed opposite each other, a duplexer, two switches and coax cabling.

\item {\bf Propulsion:}	The propulsion subsystem for SA-200S bus will be used as is. This will ensure a minimal amount of engineering required. System is a blowdown monopropellant system
 with eight 5N thrusters, two 32 cm tanks all with 22kg propellant capacity. The system exists and was tested in may applications.
\end{itemize}

\subsection{ Interferometer on the ISS}
\label{sec:interf-ISS}

\begin{figure}[h!]
 \begin{center}
\noindent  \vskip -5pt   
\psfig{figure=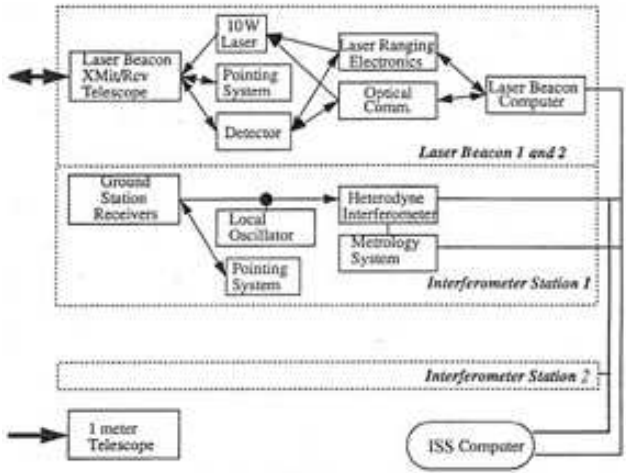,width=85mm}%,height=90mm}
\end{center}
\vskip -15pt 
  \caption{LATOR station block diagram.
 \label{fig:iss_block_giagram}}
\end{figure}

The LATOR ground station is used to interferometrically measure the angle between the two spacecraft and to transmit and receive the laser ranging signal to each of the spacecraft. A block diagram of the laser ground station is shown in Fig.~\ref{fig:iss_block_giagram} and is described in more detail below. The station on the ISS is composed of a two laser beacon stations which perform communications and laser ranging to the spacecraft and two interferometer stations which collect the downlink signal for the astrometric measurement. In addition the station uses a fiber optic link to transmit the common local oscillator to the interferometer stations.

%%************
%**************

%\subsubsection{Laser Beacon Station on the ISS and Acquisition Sequence}
\subsubsection*{Laser Beacon Station}

The laser beacon stations provide the uplink signals to the LATOR spacecraft and detect their downlink signals. The transmitter laser signal is modulated for laser ranging and to provide optical communications. Separate transmitters are used for each spacecraft each using a 1 W laser frequency doubled Nd:Yad laser at 532 nm as the source for each laser beacon. The laser beam is expanded to a diameter of 0.2 meter and is directed toward the spacecraft using a siderostat mirror. Fine pointing is accomplished with a fast steering mirror in the optical train.

During initial acquisition, the optical system of the laser beacon is modified to produce a team with a 30 arcsec divergence. This angular spread is necessary to guarantee a link with the spacecraft, albeit a weak one, in the presence of pointing uncertainties. After the acquisition sequence is complete, the beam is narrowed to a diffraction limited beam, thereby increasing the signal strength.

The downlink laser signal at 1.3 $\mu$m, is detected using a $12\times 12$ ($10 \times 10$ arcsec) array of Germanium detectors. In order to suppress the solar background, the signal is heterodyned with a local oscillator and detected within a narrow 5 kHz bandwidth. In the initial acquisition mode, the detection system searches over a 300 MHz bandwidth and uses a spiral search over a 30 arcsec angular field to find the downlink signal. Upon acquisition, the search bandwidth is decreased to 5 kHz and a quad-cell subarray is used to point the siderostat and fast steering mirrors of the beacon.

\subsubsection*{Interferometer Stations}

The interferometer stations collect the laser signal from both spacecraft to perform the heterodyne measurements needed for the interferometric angle measurement. There are a total of five receivers to make the four angular measurements needed to resolve fringe ambiguity.

\begin{figure}[h!]
 \begin{center}
\noindent  \vskip -5pt   
\psfig{figure=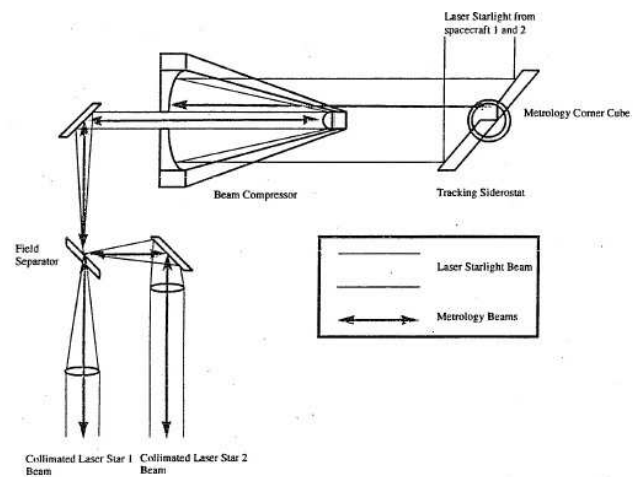,width=85mm}%,height=90mm}
\end{center}
\vskip -15pt 
  \caption{Dual feed optical system.
 \label{fig:dual-feed}}
\end{figure}

The detection and tracking system is basically similar to the receiver arm of the laser beacon described in the previous section. Light is collected by a 0.2~meter siderostat mirror and compressed with a telescope to a manageable beam size. The light from each of the spacecraft is separated using a dual feed optical system as shown in Fig.~\ref{fig:dual-feed}. A fast steering mirror is used for high bandwidth pointing of the receiver. Each spacecraft signal is interfered with a local oscillator and the phase measurement time tagged and recorded. A $6\times6$~Ge array ($5\times5$ arcsec FOV) is used to provide heterodyne acquisition and tracking of the LATOR spacecraft.

%%************
\begin{figure}[h!]
 \begin{center}
\noindent  \vskip -5pt   
\psfig{figure=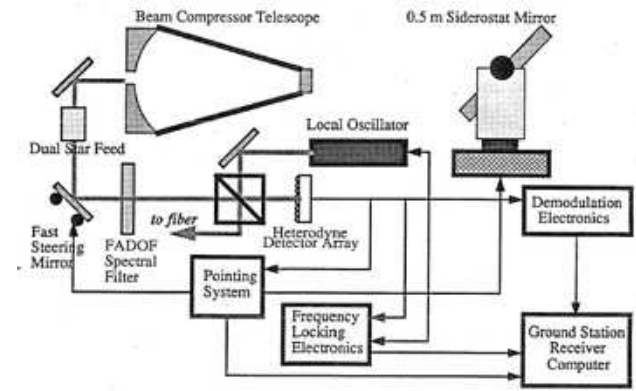,width=85mm}%,height=90mm}
\end{center}
\vskip -15pt 
  \caption{Receiver on the ISS.
 \label{fig:iss-receiver}}
\end{figure} 
%**************
%%************
%**************

\subsubsection*{ISS-Based Interferometer}

\begin{figure*}[t!]
 \begin{center}
\epsfig{figure=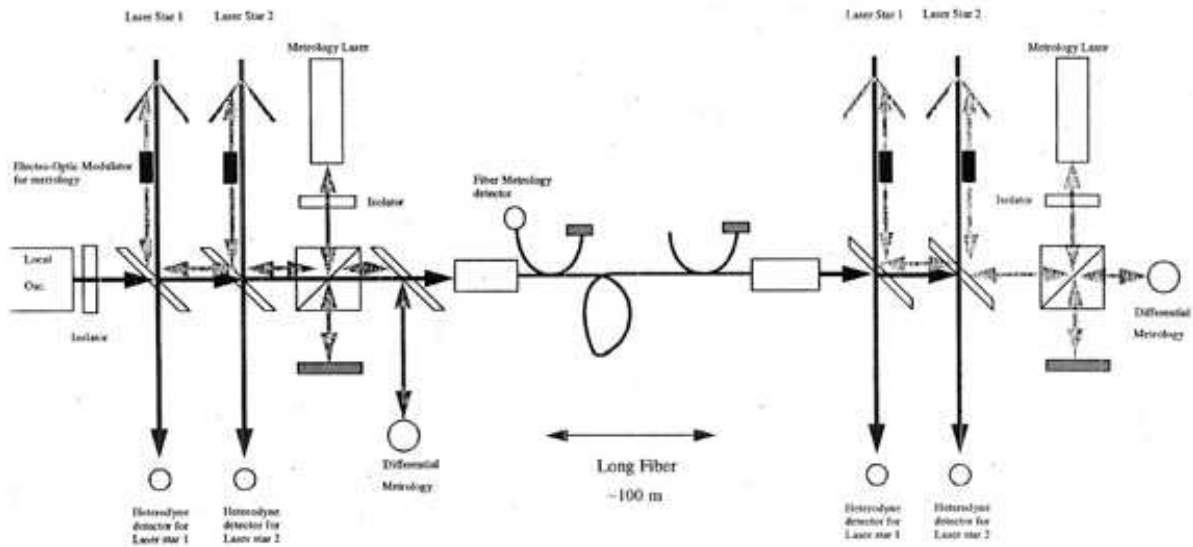,width=158mm}%,height=90mm}
\end{center}
\vskip -10pt 
  \caption{ISS-based interferometer. 
 \label{fig:iss_interf}}
\end{figure*}

Figure \ref{fig:iss_interf} shows a schematic of the ISS-based fiber interferometer used to perform the angular measurement between the spacecraft. A detailed description of how this interferometer makes its measurement is presented in Section \ref{sec:heterodine}. The interferometer includes tie heterodyne detection of the downlink signals which have been described in the previous section. The local oscillator (LO) is generated in one of the ground station receivers and is frequency locked to the laser signal from one of the spacecraft. The LO is then broadcast to the other station on the ISS through a 100 m single mode polarization preserving fiber. The heterodyne signals from all the stations (2 stations, 2 signals each) are recorded and time tagged.

%************
%**************
%%************
\begin{figure}[h!]
 \begin{center}
\noindent  \vskip -5pt   
\psfig{figure=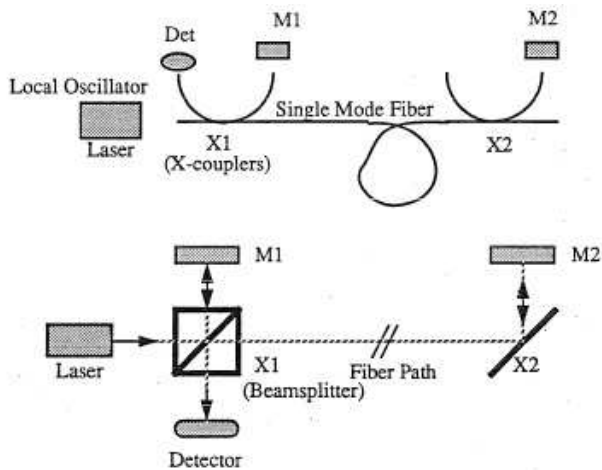,width=85mm}%,height=90mm}
\end{center}
\vskip -15pt 
  \caption{Fiber metrology system.
 \label{fig:fiber_metr}}
\end{figure} 
%**************

Figure \ref{fig:iss_interf} also shows two metrology systems used in the interferometer. The first metrology system measures the difference in optical path between the two laser signal paths and is essential to proper processing of the heterodyne data. The second metrology system measures the changes in the optical path through the fiber. This measurement monitors the length of the fiber and is used in the post processing of the interferometer data. The internal path metrology system, shown in the figure, measures the paths from corner cube on the siderostat mirror (shown as two, really only one) to the metrology beam sputter. It is essential that the laser metrology system be boresighted to the laser signal path so the correct distance is measured. A Michelson interferometer with a frequency shift in one arm measure changes in the length of each signal path. Both spacecraft signal paths are measured simultaneously. This is accomplished by using an electro-optic cell and modulating each beam at a different frequency. A He-Ne laser is used as the light source for this metrology system. Filters at the output of the detector are then used to separate the signals corresponding to each metrology beam.

The fiber metrology system measures changes in the optical path through the fiber. This system uses local oscillator signal in a Michelson configuration. Figure \ref{fig:fiber_metr} shows the correspondence between a standard Michelson interferometer and the fiber metrology system. The two X couplers serve as the beam splitters. Reflectors at the ends of the fiber couplers serve as the reference and signal mirrors. One of these reflectors is dithered to frequency shift the output signal. The phase measurement at the detector measures changes in the path length between points X1 and X2, if Ml-X1 and M2-X2 are held constant. This is accomplished by placing the X couplers and mirrors at each end of the fiber on a single thermally stable optical breadboard.

%************
\begin{figure*}[t!]
 \begin{center}
\epsfig{figure=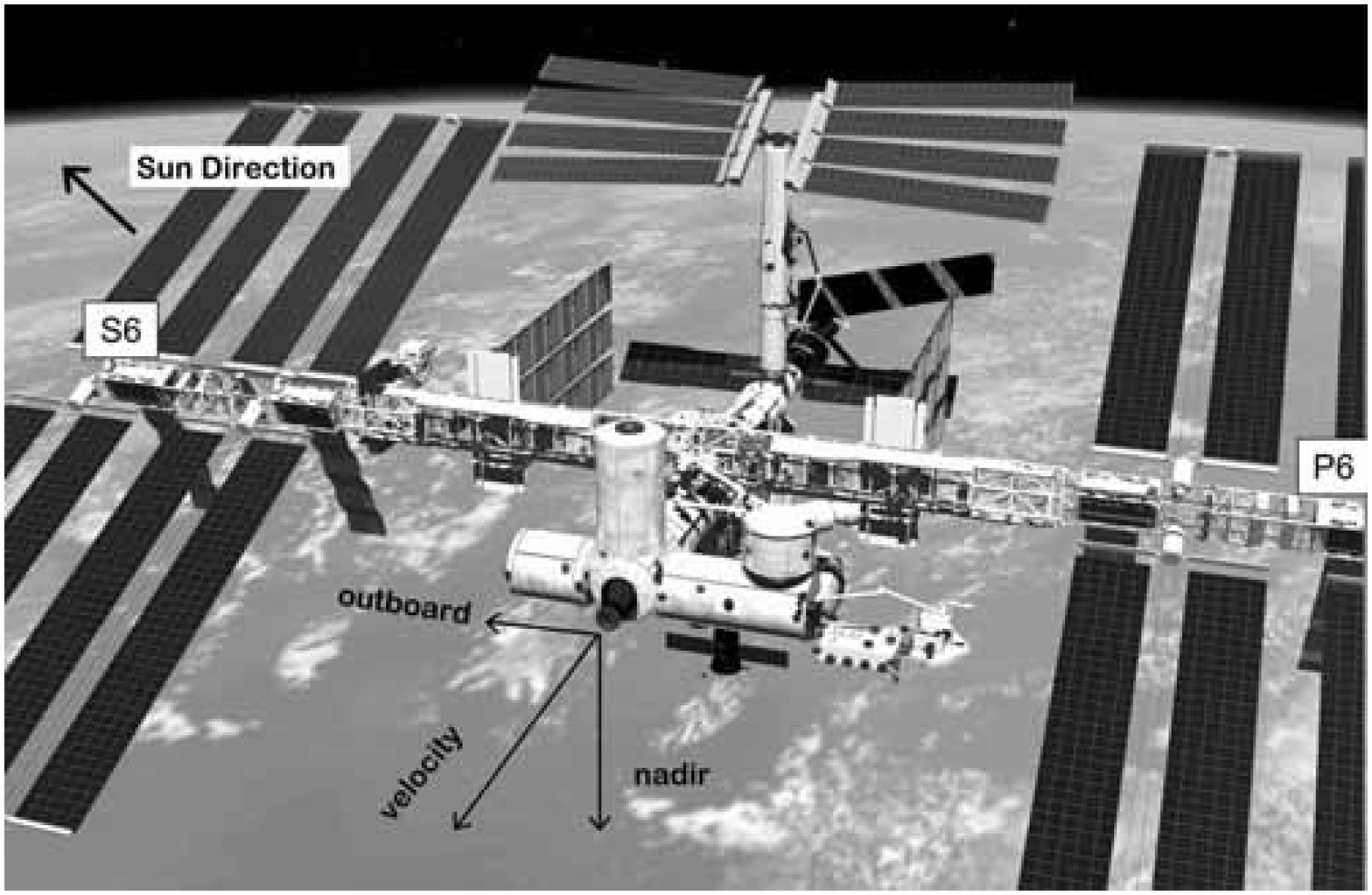,width=158mm}%,height=90mm}
\end{center}
\vskip -10pt 
  \caption{Location of the LATOR interferometer on the ISS. To utilize the inherent ISS sun-tracking capability, the LATOR optical packages will be located on the outboard truss segments P6 and S6 outwards. 
 \label{fig:iss_config}}
\end{figure*} 
%**************

The interferometer will be formed by the two optical transponder assemblies with dimensions of approximately 
0.6 m $\times$ 0.6 m $\times$ 0.6 m for each assembly (Fig.~\ref{fig:iss_config}). The mass of each telescope assembly will be about 120 kg. The location of these packages on the ISS and their integration with the ISS's power, communication and attitude control system are given below:

\begin{itemize}
\item Two LATOR transponders will be physically located and integrated with the ISS infrastructure. The location will enable the straight-line separation between the two transponders of $\sim$100 m and will provide a clear line-of-site (LOS) path between the two transponders during the observation periods. Both transponder packages will have clear LOS to their corresponding heliocentric spacecraft during pre-defined measurement periods.   

\item The transponders will be physically located on the ISS structure to maximize the inherent ISS sun-tracking capability.  The transponders will need to point towards the Sun during each observing period.  By locating these payloads on the ISS outboard truss segments (P6 and S6 outwards), a limited degree of automatic sun-tracking capability is afforded by the alpha-gimbals on the ISS.

\item The minimum unobstructed LOS time duration between each transponder on the ISS and the transponders and their respective spacecraft will be 58 minutes per the 92 min orbit of the ISS.   
 
\item	The pointing error of each transponder to its corresponding spacecraft will be no greater than 1 $\mu$rad for control,  1 $\mu$rad for knowledge, with a stability of 0.1 $\mu$rad/sec, provided by combination of the standard GPS link available on the ISS and $\mu$-g accelerometers. 

\end{itemize}

\subsection{ LATOR Instrument}

\begin{figure}[h!]
 \begin{center}
\noindent  \vskip -5pt   
\psfig{figure=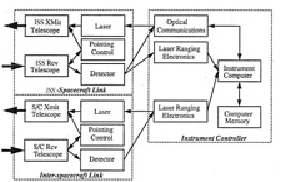,width=85mm}%,height=90mm}
\end{center}
\vskip -15pt 
  \caption{LATOR instrument subsystem block diagram.
 \label{fig:instrument}}
\end{figure}

The LATOR instrument is used to perform laser ranging between the two spacecraft; it is also used (the second set) for laser ranging and optical communications between the spacecraft and the ISS.  
Figure \ref{fig:instrument} shows a block diagram of the instrument subsystems which are describe in more detail below.

%%************
%**************
%%************
\begin{figure}[h!]
 \begin{center}
\noindent  \vskip -5pt   
\psfig{figure=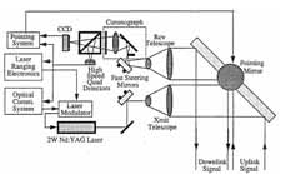,width=85mm}%,height=90mm}
\end{center}
\vskip -15pt 
  \caption{S/C Transmitter \& Receiver (ISS-Space Link).
 \label{fig:transm_receiver}}
\end{figure} 
%**************

\begin{itemize}
\item {\bf ISS-SC Receiver/Transmitter:} The ISS/SC receiver performs the acquisition, tracking, and detection of the signals from the ISS (Figure \ref{fig:transm_receiver}). This uplinked signal will be sent at 1064 nm and will contain modulation to perform both laser ranging and to send control signals to the spacecraft. The signals from the ISS are detected by a telescope with a collecting aperture of 20 cm. A coronograph will by used to suppress stray light from the Sun. In addition a combination of a wideband interference filter and a narrow band Faraday Anomalous Dispersion Optical Filter (FADOF) will used to reject light outside a 0.005 nm band around the laser line. The incoming signal is subdivided with one portion going to a high bandwidth detector and the other to an acquisition and tracking CCD array. Using a $64 \times 64$ CCD array with pixels sized to a diffraction limited spot, this array will have a 5 arcmin field of view which is greater than the pointing knowledge of the attitude control system and the point ahead angle (40 arcsec). After acquisition of the ISS beacon, a $2\times 2$ element subarray of the CCD will be used as a quad cell to control the ISS-SC two axis steering mirror. This pointing mirror is common to both the receiver and transmitter channel to minimize misalignments between the two optical systems due to thermal variations. The pointing mirror will have 10 arcminute throw and a pointing accuracy of 0.5 arcsec which will enable placement of the uplink signal on the high bandwidth detector.

%%************
\begin{figure}[h!]
 \begin{center}
\noindent  \vskip -5pt   
\psfig{figure=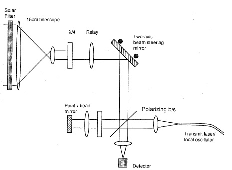,width=85mm}%,height=90mm}
\end{center}
\vskip -15pt 
  \caption{ISS-S/C Link with common optics: spiral scanning spatial acquisition; open loop point ahead control with piezo actuators; fiber-coupled, frequency stabilized transmitter; pupil planes at the steering mirror and mixing apertures.
 \label{fig:optics}}
\end{figure} 
%**************

The ISS-SC transmitter sends a laser signal to both the interferometer collectors and the beacon receivers. The signal will be encoded for both ranging and communication information. In particular, the transmitted signal will include the inter-spacecraft ranging measurements. The transmitter uses a 2~W frequency stabilized Nd:YAG laser at 1.3 $\mu$m. A 5~kHz line width is required to simplify heterodyne detection at the ground station. A 0.15 meter telescope is used to transmit the laser beam and a steering mirror is used for pointing. The mirror uses information from the attitude control system (ACS), the quad-cell detector in the receiver, and the point ahead information from the instrument controller to determine the transmit direction. A fast steering mirror is used to maintain high bandwidth pointing control for both the transmitter and receiver. 

We have also considering the possibility of using a common optical system for both the transmitter and receiver. Figure \ref{fig:optics} shows a schematic of such a transmitter/receiver system. Because of the difference in the receive and transmit wavelengths, dichroic beam splitters and filters are used to minimize losses from the optics and leakage into the detectors. In this scheme a point ahead mirror is used maintain a constant angular offset between the received and transmitted beams. Because of the common optical elements, this system is more tolerant to misalignments than the previous configuration.

\item {\bf Inter-S/C Receiver/Transmitter:} The inter-S/C receiver/transmitter uses two separate optical systems. The receiver detects the laser ranging signal from the other spacecraft. The receiver is similar in design to the ISS-S/C receiver subsystem. Since there is no solar background contribution, the coronograph and FADOF filter have been removed. Detection of the signal is accomplished using a CCD for acquisition and a quad cell subarray for tracking. The tracking signal is also used to control the pointing of the transmitter minor. A separate high bandwidth detector is used for detecting the laser ranging signal.

%%************
\begin{figure}[h!]
 \begin{center}
\noindent  \vskip -5pt   
\psfig{figure=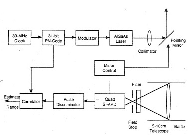,width=85mm}%,height=90mm}
\end{center}
\vskip -15pt 
  \caption{ISS-S/C Link with common optics: spiral scanning spatial acquisition; open loop point ahead control with piezo actuators; fiber-coupled, frequency stabilized transmitter; pupil planes at the steering mirror and mixing apertures.
 \label{fig:inter_sc}}
\end{figure} 
%**************

The inter-S/C transmitter sends the laser ranging signal to the other spacecraft. The transmitter uses a 780 nm laser with an output power of 10 mW. The transmitter and receiver telescopes have an aperture of 0.1 m diameter. Because of the close proximity of the LATOR spacecraft, thermal drifts which cause misalignments between the transmitter and receiver optical systems can be sensed and corrected rapidly. In addition, the LATOR geometry requires minimal point ahead since the transverse velocity between spacecraft is nearly zero.

\item {\bf Instrument Controller:} The instrument controller subsystem contains the remainder of the instrument hardware. This includes the electronics needed for the laser ranging and optical communications as well as the computer used to control the instrument. The instrument computer will take information from the attitude control system and receiver subsystems in order to control the pointing of the transmit subsystems and the modulation of their laser signals.
\end{itemize}

%\subsubsection{LATOR Instrument}

The LATOR instrument in each of the two spacecraft consist of three laser metrology transmitters and receivers which can be gimbaled to point at the other spacecraft, and a camera system to acquire the incoming laser signals and to control the pointing of the outgoing beams.  In addition, the instrument contains a laser ranging transponder in order to determine the spacecraft position from the ground.   
\begin{itemize}
\item{\bf Laser Metrology Transceiver Subsystem}
The metrology transceiver consists of the laser, modulators, optics, and frequency stabilizer.  The laser light is first frequency stabilized to better than 1 part in 10$^{10}$, in order to make the measurements.  The laser light is then frequency modulated in order to produce the heterodyne signal and distinguish between incoming and outgoing beams.  Finally light is then collimated and injected into the beam launcher optics. The incoming metrology signal is received by the beam launcher optics and is interfered with the local laser.  A cat's eye retroreflector serves as the spacecraft fiducial and is common to all three beam launchers.  
  
\item{\bf Beam Launcher/Receiver Optics: }
In the current instrument design, the modulated laser beam is injected using a polarization preserving single mode fiber and expanded to a 0.5 cm beam.  A cat's eye retroreflector is one of several devices that can be used as the metrology fiducial and is common to the three metrology beams. The cat's eye uses two optically contacted concentric hemispheres with radius of $\sim$ 10 cm and $\sim$ 20 cm. The cat's eye is sized many times larger than the beam in order to minimize the effect of spherical aberration. The beam is then expanded to a 5 cm beam using a refractive telescope. A refractive design was chosen because changes in the optical path are relatively insensitive to changes in the position and orientation of the optical elements.  

\item{\bf Acquisition Camera (AC) Subsystem: }
AC that will be used as the sensor for pointing the metrology beam.  In the previous study the $512\times512$ camera was used to detect the position of the incoming laser beam to 0.5 arcsec over a 1 degree field by interpolating the centroid of the spot to 0.1 pixel. Three cameras were used to track each of the incoming metrology beams. The outgoing laser beam was retroreflected from the alignment corner cube to produce a spot on the acquisition camera on which to servo the pointing gimbal.  

\item{\bf Pointing Subsystem:} In the current instrument design the entire beam launcher optical assembly is gimbaled to point the metrology beam to the target spacecraft.  The 2-axis gimbal has a center of rotation at the center of the cat's eye retro reflector.  This optical arrangement measures the distance between the optical fiducials and is not sensitive to slight misalignments to first order.  The gimbal will have a range of 1 degree and a pointing resolution of 0.5 arcsec. 
\end{itemize}

\subsection{LATOR Operations}
\label{sec:operations}

This section describes the sequence of events that lead to the signal acquisition and that occurs during each observation period.  This sequence will be initiated at the beginning of the experiment period, after ISS emergence from the Earth's shadow. It assumed that boresighting of the spacecraft attitude with the spacecraft transmitters and receivers have already been accomplished. This sequence of operations is focused on establishing the ISS to spacecraft link. The interspacecraft link is assumed to be continuously established after final-deployment (at $\sim15^\circ$ off the Sun), since the spacecraft never lose line of sight with one another. 

The laser beacon transmitter at the ISS is expanded to have a beam divergence of 30 arcsec in order to guarantee illumination of the LATOR spacecraft. After re-emerging from the Earth's shadow this beam is transmitted to the craft and reaches them in about 18 minutes. At this point, the LATOR spacecraft acquire the expanded laser beacon signal. In this mode, a signal-to-noise ratio (SNR) of 4 can be achieved with 30 seconds of integration. With an attitude knowledge of 10 arcsec and an array field of view of 30 arcsec no spiral search is necessary. Upon acquisition, the receiver mirror on the spacecraft will center the signal and use only the center quad array for pointing control. Transition from acquisition to tracking should take about 1 minute. Due to the weak uplink intensity, at this point, tracking of the ISS station is done at a very low bandwidth. The pointing information is fed-forward to the spacecraft transmitter pointing system and the transmitter is turned on. The signal is then re-transmitted down to the ISS with a light-travel time of 18 minutes.

Each interferometer station and laser beacon station searches for the spacecraft laser signal. The return is heterodyned with using an expanded bandwidth of 300~MHz. In this case, the solar background is the dominant source of noise, and an SNR ratio of 5 is achieved with 1 second integration. Because of the small field of view of the array, a spiral search will take 30 seconds to cover a 30 arcsec field. Upon acquisition, the signal will be centered on the quad cell portion of the array and the local oscillator frequency locked to the spacecraft signal. The frequency band will then be narrowed to 5 kHz. In this regime, the solar background is no longer the dominant noise source and an SNR of 17.6 can be achieved in only 10 msec of integration. This will allow one to have a closed loop pointing bandwidth of greater than 100 Hz and be able to compensate for the tilt errors introduced by the atmosphere. The laser beacon transmitter will then narrow its beam to be diffraction limited ($\sim$1 arcsec) aid point toward the LATOR spacecraft. This completes the signal acquisition phase, the entire architecture is in-lock and transmits scientific signal.  This procedure is re-established during each 92-minute orbit of the ISS.

In the next section we will consider the LATOR preliminary astrometric error budget. 

\section{Astrometric Performance}
\label{sec:error_bud}

In our design considerations we address two types of instrumental errors, namely the offset and scale errors. Thus, in some cases, when a measured value has a systematic offset of a few pm, there are may be instrumental errors that lead to further offset errors.  There are many sources of offset (additive)  errors caused by imperfect optics or imperfectly aligned optics at the pm level; there also many sources for scale errors. We take a comfort in the fact that, for the space-based stellar interferometry, we have an ongoing technology program at JPL;  not only this program has already demonstrated metrology accurate to a sub-pm level, but has also identified a number of the error sources and developed methods to either eliminate them or to minimize their effect at the required level.

The second type of error is a scale error. For instance, in order to measure $\gamma$ to one part in  $10^{8}$ the laser frequency also must be stable to at least to $10^{-8}$ long term; the lower accuracy would result in a scale error. The measurement strategy adopted for LATOR would require the laser stability to only $\sim$1\% to achieve accuracy needed to measure the second order gravity effect. Absolute laser frequency must be known to $10^{-9}$ in order for the scaling error to be negligible. Similarly robust solutions were developed to address the effects of other known sources of scale errors. 

There is a considerable effort currently underway at JPL to evaluate a number of potential errors sources for the LATOR mission, to understand their properties and establish methods to mitigate their contributions. (A careful strategy is needed to isolate the instrumental effects of the second order of smallness; however, our experience with SIM \cite{mct, Turyshev01_1, Turyshev01_2} is critical in helping us to properly capture their contribution in the instrument models.)  The work is ongoing, this is why the discussion below serves for illustration purposes only. We intend to publish the corresponding analysis and simulations in the subsequent publications.

%********************Start Table******************
\begin{table*}[ht!]
\caption{LATOR Mission Summary: Science Objectives 
\label{tab:summ_science}}
\vskip 5pt
\begin{tabular}{m{12.1cm}} \hline \hline 

\reff\hskip4pt$\bullet$\hskip6pt
To search for cosmological remnants of scalar field in the solar system.

\reff\hskip4pt$\bullet$\hskip6pt
To access the most intense gravitational environment in the solar system and test a number of dynamical scenarios in this new field-strength  regime.

\reff\hskip4pt$\bullet$\hskip6pt
To measure the key Eddington parameter $\gamma$ with  accuracy of 1 part in 10$^{8}$, a factor of 3,000 improvement in the tests of gravitational light deflection.

\reff\hskip4pt$\bullet$\hskip6pt
To directly measure the PPN parameter $\beta$ to $\sim1$\% accuracy

\reff\hskip4pt$\bullet$\hskip6pt
To measure effect of the second order light deflection with accuracy of $\sim1\times 10^{-3}$, including first ever measurement of the PPN parameter $\delta$.  

\reff\hskip4pt$\bullet$\hskip6pt
To measure the solar quadrupole moment (using the theoretical value of the solar quadrupole moment $J_2\simeq10^{-7}$) to 1 part in 20.

\reff\hskip4pt$\bullet$\hskip6pt
To directly measure the frame dragging effect on the light with $\sim 1\times10^{-2}$ accuracy.
\\[2pt]\hline 
\end{tabular}
\end{table*}
 
%********************End Table******************

\subsection{Optical Performance}
 
The laser interferometers use $\sim$2W lasers and $\sim$20 cm optics for transmitting the light between spacecraft. Solid state lasers with single frequency operation are readily available and are relatively inexpensive.   For SNR purposes we assume the lasers are ideal monochromatic sources (with $\lambda = 1.3~ \mu$m). For simplicity we assume the lengths being measured are 2AU = $3\times 10^8$ km. The beam spread is estimated as $\sim 1~\mu$m/20~cm = 5 $\mu$rad (1 arcsec). The beam at the receiver is $\sim$1,500 km in diameter, a 20 cm receiver will detect $1.71 \times 10^2$ photons/s assuming 50\% q.e. detectors. Given the properties of the CCD array it takes about 10 s to reach the desirable SNR of $\sim2000$ targeted for the detection of the second order effects. In other words, a 5 pm resolution needed for a measurement of the PPN parameter $\gamma$ to the accuracy of one part in $\sim10^{8}$ is possible with $\approx10$~s of integration.

As a result, the LATOR experiment will be capable of measuring the angle between the two spacecraft to $\sim0.05$~prad, which allows light deflection due to gravitational effects to be measured to one part in $10^8$. Measurements with this accuracy will lead to a better understanding of gravitational and relativistic physics. 

In our analysis we have considered various potential sources of systematic error.  
This information translates to the expected accuracy of determination of the differential interferometric delay of $\sim \pm5.4$ pm, which translates in the  measurement of the PPN parameter $\gamma$ with accuracy of  
%\begin{eqnarray}
$\sigma_\gamma \sim 0.9  \times 10^{-8}.$
%\label{eq:accuracy}
%\end{eqnarray}
 This expected instrumental accuracy is clearly a very significant improvement compared to other currently available techniques. This analysis serves as the strongest experimental motivation to  conduct the LATOR experiment.

\subsection{\label{sec:expect_accuracy}Expected Measurement Accuracy}

Here we summarize our estimates of the expected accuracy in measurement of the relativistic parameters of interest.
The first order effect of light deflection in the solar gravity caused by the solar mass monopole is $\alpha_1=1.75$ arcsec; this value corresponds to an interferometric delay of $d\simeq b\alpha_1\approx0.85$~mm on a $b=100$~m baseline. Using laser interferometry, we currently able to measure  distances with an accuracy (not just precision but accuracy) of $\leq$~1~pm. In principle, the 0.85 mm gravitational delay can be measured with $10^{-9}$ accuracy versus $10^{-5}$ available with current techniques. However, we use a conservative estimate  of 10 pm for the accuracy of the delay which would lead to a single measurement of $\gamma$ accurate to 1 part in $10^{8}$ (rather than 1 part in $10^{9}$), which would be already a factor of 3,000 accuracy improvement when compared to the recent Cassini result \cite{cassini_ber}. 

Furthermore, we have targeted an overall measurement accuracy of 10 pm per measurement, which for $b=100$~m this translates to the accuracy of 0.1 prad $\simeq 0.02 ~\mu$as. With 4 measurements per observation, this yields an accuracy of $\sim5.8\times 10^{-9}$ for the first order term.
The second order light deflection is approximately 1700 pm and with 10 pm accuracy and the adopted measurement strategy it could be measured with accuracy of $\sim2\times 10^{-3}$, including first ever measurement of the PPN parameter $\delta$.  The frame dragging effect would be measured with $\sim 1\times10^{-2}$ accuracy and the solar quadrupole moment (using the theoretical value of the solar quadrupole moment $J_2\simeq10^{-7}$) can be modestly measured to 1 part in 20, all with respectable SNRs (see Table \ref{tab:summ_science}).

The final error would have contributions from multiple measurements of the light triangle's four attributes (to enable the redundancy) taken by range and interferometer observations at a series of times. The corresponding errors will be combined with those from orbital position and velocity coordinate uncertainty.  These issues are currently being investigated in the mission covariance analysis; the detailed results of this analysis will be reported elsewhere. However, our current understanding of the  expected mission and instrumental accuracies suggests that LATOR will offer a very significant improvement compare to any other available techniques. This conclusion serves as the strongest experimental motivation to conduct the LATOR experiment.  

%************************************************
\section{Conclusions}
\label{sec:conc}

Concluding, we would like to summarize the most significant results of our LATOR mission study. The most natural question is ``Why is LATOR potentially orders of magnitude more sensitive and less expensive?''

First of all, there is a significant advantage in using optical wavelengths as opposed to the microwaves -- the present navigational standard. This is based on the fact that solar plasma effects decrease as $\lambda^2$ and, in the case of LATOR, we gain a factor of $10^{10}$ reduction in the solar plasma optical path fluctuations by simply moving from $\lambda=10$~cm to $\lambda=1 ~\mu$m. Another LATOR's advantage is its independence of a drag-free technology. In addition, the use of a redundant optical truss offers an excellent alternative to an ultra-precise orbit determination. This feature also makes LATOR insensitive to spacecraft buffeting from solar wind and solar radiation pressure.

Furthermore, the use of existing technologies, laser components and spacecraft make this mission a low cost experiment. Thus, 1~W lasers with sufficient frequency stability and $>10$ years lifetime already developed for optical telecom and also are flight qualified for SIM. Additionally, small optical apertures $\sim$10-20cm are sufficient and provide this experiment with a high signal-to-noise ratio. There also a significant advantage in using components with no motorized moving parts.
This all makes LATOR an excellent candidate for the next flight experiment in fundamental physics. Table \ref{tab:summ_science} summarizes the science objectives for this mission. 

The LATOR mission aims to carry out a test of the curvature of the solar system's gravity  field with an accuracy better than 1 part in 10$^{8}$. In spite of the previous space missions exploiting radio waves for tracking the spacecraft, this mission manifests an actual breakthrough in the relativistic gravity experiments as it allows to take full advantage of the optical techniques that recently became available.  The LATOR experiment has a number of advantages over techniques that use radio waves to measure gravitational light deflection. The optical technologies allows low bandwidth telecommunications with the LATOR spacecraft. The use of the monochromatic light enables the observation of the spacecraft almost at the limb of the Sun. The use of narrowband filters, coronagraph optics and heterodyne detection will suppress background light to a level where the solar background is no longer the dominant noise source. The short wavelength allows much more efficient links with smaller apertures, thereby eliminating the need for a deployable antenna. Finally, the use of the ISS enables the test above the Earth's atmosphere -- the major source of astrometric noise for any ground based interferometer. This fact justifies LATOR as a space mission.

The LATOR mission will utilize several technology solutions that recently became available. In particular, signal acquisition on the solar background will be done with a full-aperture narrow band-pass filer and coronagraph. The issue of the extended structure vibrations of the will be addressed by using $\mu$-g accelerometers. (The use of the accelerometers was first devised for SIM, but at the end their utilization is not needed. The Keck Interferometer uses accelerometers extensively.) Finally, the problem of monochromatic fringe ambiguity that complicated the design of the previous version of the experiment \cite{yu94} and led to the use of variable baselines lengths -- is not an issue for LATOR. This is because the orbital motion of the ISS provides variable baseline projection that eliminates this problem for LATOR.  

The LATOR experiment technologically is a very sound concept; all technologies that are needed for its success have been already demonstrated as a part of the JPL's interferometry program.  The LATOR experiment does not need a drag-free system, but uses a geometric redundant optical truss to achieve a very precise determination of the interplanetary distances between the two micro-spacecraft and a beacon station on the ISS. The interest of the approach is to take advantage of the existing space-qualified optical technologies leading to an outstanding performance in a reasonable mission development time.  The  availability of the ISS makes this mission concept realizable in the very near future; the current mission concept calls for a launch as early as in 2011 at a cost of a NASA MIDEX mission.   

This mission may become a 21st century version of Michelson-Morley experiment in the search for a  cosmologically evolved scalar field in the solar system. As such, LATOR will lead to very robust advances in the tests of fundamental physics: it could discover a violation or extension of GR, or reveal the presence of an additional long range interaction in the physical law.  There are no analogs to the LATOR experiment; it is unique and is a natural culmination of solar system gravity experiments. 

\vfill
%**************************************
\acknowledgments
The work described here was carried out at the Jet Propulsion Laboratory, California Institute of Technology, under a contract with the National Aeronautics and Space Administration.

%**************************************


\begin{thebibliography}{99}

\bibitem{viking_shapiro1} 
	I.I. Shapiro, C.C.Counselman, III, and R.W. King, 
%``Verification of the Principle of Equivalence for Massive Bodies,'' 
Phys. Rev. Lett. {\bf 36}, 555 (1976).

\bibitem{viking_reasen}	
	R.D. Reasenberg  et al., 
%``Viking relativity experiment: Verification of signal retardation by solar gravity,'' 
ApJ Lett. {\bf 234}, L219 (1979).

\bibitem{viking_shapiro2} 
	I.I. Shapiro, et al.,
% ``The Viking relativity experiment,'' 
JGR {\bf 82}, 4329 (1977).

\bibitem{anderson02}  
J.~D. Anderson {et al}.,
%,  Lau, E. L., Turyshev, S. G.,  Williams, J. G.,  Nieto, M. M., ``Recent Results for Solar-System Tests of General Relativity.'' Presented at 200-th AAS Meeting, Albuquerque, NM (2-6 June 2002).  Paper \#12.06.  
BAAS {\bf 34}, 833 (2002).

\bibitem{RoberstonCarter91}	
	D.S. Robertson, W.E. Carter and W.H. Dillinger, 
Nature {\bf 349} 768 (1991).

\bibitem{Lebach95}	
	D.E. Lebach et al., 
%, Corey, B. E., Shapiro, I. I., Ratner, M. I., Webber, J. C., Rogers, A. E. E., Davis, J. L. \& Herring, T. A., 
Phys. Rev. Lett. {\bf 75}, 1439 (1995). 

\bibitem{Shapiro_SS_etal_2004}
S.~S. Shapiro, {et al.}, 
%J.~L. Davis, D.~E. Lebach, and J.~S. Gregory,
%''Measurement of the Solar Gravitational Deflection of RadioWaves using Geodetic Very-Long-Baseline Interferometry Data, 1979–1999''
Phys. Rev. Lett. {\bf 92}, 121101 (2004).

\bibitem{Ken_LLR68}	
	K. Nordtvedt, Jr., 
%``Testing Relativity with Laser Ranging to the Moon,'' 
Phys. Rev. {\bf 170}, 1186 (1968).

\bibitem{Ken_LLR91}	
	K. Nordtvedt, Jr.,  
%``Lunar Laser Ranging Re-examined: The Non-Null Relativistic Contribution,'' 
Phys. Rev. D {\bf 43}, 10 (1991).

%\bibitem{Ken_LLR98}	
%	K. Nordtvedt, Jr.,  
%%Nordtvedt, K., Jr., ``Optimizing the observation schedule for tests of gravity in lunar laser ranging and similar experiments,'' Class. Quantum Grav. 
%CQG {\bf 15}, 3363 (1998).

\bibitem{Ken_LLR30years99} 
	K. Nordtvedt, Jr.,  
%Nordtvedt, K., ``30 years of lunar laser ranging and the gravitational interaction,'' 
CQG {\bf 16},  A101 (1999).

\bibitem{Ken_LLR_PPNprobe03}	
	K. Nordtvedt, Jr.,   
%Nordtvedt, K., Jr., ``Lunar Laser Ranging - A Comprehensive Probe of Post-Newtonian Gravity'', 
[gr-qc/0301024].

\bibitem{JimSkipJean96}	
	J.~G. Williams, X.~X. Newhall, and J.~O. Dickey, 
% `Relativity Parameters Determined from Lunar Laser Ranging,'' Phys. Rev. D 
Phys. Rev. D {\bf 53}, 6730 (1996).

\bibitem{Williams_etal_2001}	
	J.~G. Williams  et al., 
%, Anderson, J. D., Boggs, D. H., Lau, E. L., \& Dickey, J. O., ``Solar System Tests for Changing Gravity,'' AAS Meeting, Pasadena, CA, June 3-7, 2001, 
BAAS {\bf 33}, 836 (2001).

\bibitem{LLR_beta_2004} J.~G. Williams, S.~G. Turyshev, D.~H. Boggs, %2004, {\it New Results in LLR Tests of Relativistic Gravity}.
      To be published {Phys. Rev. Lett.}, 2004.  

%\bibitem{[28]}	
%	L. Iess et al., 
%%, Giampieri, G., Anderson, J. D., \&  Bertotti, B., Class. Quant. Gr. 
%CQG {\bf 16}, 1487 (1999).

\bibitem{cassini_ber} 
	B. Bertotti, L. Iess, and P. Tortora, 
%``A test of general relativity using radio links with the Cassini spacecraft'',   
Nature {\bf 425}, 374  (2003).

%\bibitem{damour_nordtvedt1} 
\bibitem{damour_nordtvedt} 
	T. Damour, K. Nordtvedt, Jr., 
%``General Relativity as a Cosmological Attractor of Tensor Scalar Theories'', Phys. Rev. Lett., 
Phys. Rev. Lett. {\bf 70}, 2217 (1993);
%
%\bibitem{damour_nordtvedt2}	
%	T. Damour and K. Nordtvedt, Jr., 
%Damour, T. \& Nordtvedt, K. Jr., ``Tensor-scalar cosmological models and their relaxation toward general relativity,'' Phys. Rev. 
Phys. Rev. D {\bf 48}, 3436  (1993).


%\bibitem{DamourPolyakov94_1}
\bibitem{DamourPolyakov94}	
	T. Damour, A.~M. Polyakov, 
%, ``String Theory and Gravity,'' Gen. Relativ. Gravit., 
 GRG  {\bf 26}, 1171  (1994);
%
%\bibitem{DamourPolyakov94_2}	
%	T. Damour, A.M. Polyakov, 
%``The string dilaton and a least coupling principle,'' 
 Nucl. Phys.  B {\bf 423}, 532  (1994).

\bibitem{DPV02} 
	T. Damour, F. Piazza, and G. Veneziano,
%``Runaway dilaton and equivalence principle violations''
Phys. Rev. Lett. {\bf 89}, 081601, (2002) [gr-qc/0204094];
%
%\bibitem{DPV02b} 
%T. Damour, F. Piazza and G. Veneziano,
%``Violations of the equivalence principle in a dilaton-runaway scenario''
Phys. Rev. D {\bf 66}, 046007, (2002) [hep-th/0205111].

%\bibitem{Damour_EFarese96a}
\bibitem{Damour_EFarese96} 
T. Damour, G. Esposito-Farese,  Phys. Rev. D {\bf 53} 5541 (1996);
%\bibitem{Damour_EFarese96b}
%T. Damour, G. Esposito-Farese,
Phys. Rev. D {\bf 54}, 1474 (1996);

\bibitem{Ken_2PPN_87}	
	K. Nordtvedt, Jr., 
%, ``Probing Gravity to the 2nd Post-Newtonian Order and to one part in 10$^7$ Using the Sun,'' 
ApJ {\bf 320}, 871 (1987).

\bibitem{lator_cqg_2004}  S.~G. Turyshev, M. Shao, and K.L. Nordtvedt, Jr.,
%{\it The Laser Astrometric Test of Relativity (LATOR) Mission}. 
CQG {\bf 21}, 2773 (2004),  gr-qc/0311020 
%/ doi:10.1088/0264-9381/21/12/001

\bibitem{footnote} A version of LATOR with a ground-based receiver was proposed in 1994 and performed under NRA 94-OSS-15 \cite{yu94}. Due to atmospheric turbulence and seismic vibrations that are not common mode to the receiver optics, a very long baseline interferometer (30 km) was proposed. This interferometer could only measure the differential light deflection to an accuracy of 0.1 $\mu$as, with a spacecraft separation of less than 1 arc minutes.

\bibitem{yu94} 
	J. Yu, et al., 
%M. Shao, Y. Gursel and R. Hellings, 
SPIE 2200, 325 (1994); 
%
%\bibitem{Shao96}	
	M. Shao  et al., 
{\it Laser Astrometric Test of Relativity (LATOR) Mission}. 
JPL Technical Memorandum  (1996).

\bibitem{teamx} 
	A. Gerber  et al.,  {\it LATOR 2003 Mission Analysis},  JPL Advanced Project Design Team (Team X) Report \#X-618 (2003).

\bibitem{mct} 
	M. Milman, J. Catanzarite, and S.~G. Turyshev, 
%%, {\it The effect of wavenumber error on the computation of path-lenght delay in white-light interferometry}.
%%     {\it Applied Optics} {\bf 41}  4884-4890 
     {Applied Optics} {\bf 41}  4884  (2002)

\bibitem{Turyshev01_1}  
	S.~G. Turyshev,   
%{\it Analytical Modeling of the White Light Fringe}. 
%     {\it Applied Optics} {\bf 42} 71-90 physics/0301026
     {Applied Optics} {\bf 42}, 71 (2003) [physics/0301026]

\bibitem{Turyshev01_2}  
	M. Milman, S.~G. Turyshev, 
%{\it Observational Model for Microarcsecond Astrometry with  the Space Interferometry Mission}. 
%{\it Optical Engeneering} {\bf 42}  1873-1883 physics/0301047
{Optical Engeneering} {\bf 42}, 1873 (2003) [physics/0301047]

\end{thebibliography}
\end{document}